\documentclass[twocolumn,eqsecnum,superscriptaddress,showpacs]{revtex4}
\usepackage{amsmath,amssymb}
\usepackage{bm}

\begin{document}
\title{Gauge non-invariance of quark-quark interactions}
\author{Takehisa \textsc{Fujita}} \email{fffujita@phys.cst.nihon-u.ac.jp}
\author{Seiji \textsc{Kanemaki}} \email{kanemaki@phys.cst.nihon-u.ac.jp}
\affiliation{Department of Physics, Faculty of Science and Technology
Nihon University, Tokyo, Japan }%
\author{Sachiko \textsc{Oshima}} 
\affiliation{Department of Physics, Faculty of Science, Tokyo Institute 
of Technology, Tokyo, Japan} 

\date{\today}

\begin{abstract}
We discuss the gauge invariance of  quark-quark potential in QCD in the perturbative 
treatment, and show that the gauge dependence of the quark-gluon interaction cannot be 
eliminated. Therefore, the quark-quark potential cannot properly be written 
in terms of physical variables in the perturbation theory. 
This is in contrast to the QED case in which the gauge invariance of 
the electron-photon interaction is guaranteed by the current conservation. As an example, 
we show that the different choices of the gauge fixing give different Coulomb interactions 
when the fermion current is not conserved. Also, we examine a possible existence of 
$QG-$state (one quark$-$one gluon) corresponding to the conserved color current, 
which should have a fractional electric charge. 

\end{abstract}

\pacs{11.10.Kk,12.38.Gc,11.15.Ha}
\maketitle

\section{ Introduction}
The strong interaction is described by Quantum Chromodynamics, and hadrons 
are described in terms of quark and gluon dynamics \cite{fay}. Quarks and gluons are 
confined inside hadrons since they have color charges. Each color current 
of quark and gluon is gauge dependent, and therefore they are not physical 
observables. The confinement of quarks and gluons is due to the nonabelian 
character of the gauge field theory, and the dynamics of quarks and gluons should 
be controlled by the kinematical constraints of the nonabelian nature 
of the gauge field theory. For example, the quark color current which is 
gauge dependent cannot come out from hadrons 
as the MIT bag model assumes \cite{MIT}. In this respect, the confinement 
should not simply be originated from the force between quarks and anti-quarks.  
In fact, it has recently been shown that Wilson's area law between two quarks \cite{wil} is 
not physically justified and his area law is the result of an artifact 
of Wilson's action. Indeed, Wilson's action inevitably picks up unphysical 
contributions to the Wilson loop integral \cite{asaga} which results in a fictitious 
area law. Therefore, there is no confinement which arises from a linear potential 
between quarks inside hadrons. In addition, it is, even more, difficult to define 
a force between quarks since the color charge of quarks is not a conserved quantity.  

The gauge dependence of the quark color charge is well known. Yet, it is somewhat 
puzzling that one often finds discussions in field theory textbooks concerning the shape 
of the quark-quark potential in the perturbative calculations. The potential between 
quarks may be defined if the color charge of quarks is a conserved quantity. 
However, since the color charge of quarks is not conserved and is indeed gauge dependent 
in QCD, it is difficult to consider any interactions 
between quarks whose color charges are arbitrary time dependent numbers. 

In this paper, we clarify that the gauge dependence of the quark-gluon interaction 
cannot be taken away by any kinds of transformation. Therefore, the  quark-quark potential 
which is derived in the second and higher order perturbative calculations 
of the quark-gluon interaction cannot be a physical quantity since it depends 
on the gauge variable $\chi^a$ which is an arbitrary function of space and time.  

The difficulty of evaluating interactions between quarks is also related to 
the non-existence of a free quark state since one cannot carry out any 
perturbative calculations which are always based on the S-matrix formulation. 
As is well known, one has to prepare free quark states as the initial 
and final states in the S-matrix evaluations. However, since there is no free quark 
state, one does not know how to proceed the S-matrix calculation in QCD as far as 
one wishes to be sufficiently careful.

For comparison, we briefly discuss the gauge dependence of the electron-photon 
interaction in the perturbation calculation. This is carefully examined in some of 
the field theory textbooks, and  the potential between electrons is well defined 
because the gauge dependence of the electron-photon interaction 
is nicely eliminated under the vector current conservation. This is closely related 
to the fact that the electron charge is indeed conserved, and one can always define 
a free electron state as a trivially simple fact. In this respect, there is 
no problem in evaluating Feynman diagrams in a perturbation theory of QED. 

Under the condition that the fermion current is not conserved, we 
show that the different choices of the gauge fixings (temporal gauge $A_0=0$ and  
Coulomb gauge $\nabla \cdot \bm{A} =0$ ) give different Coulomb interactions.   
 This is quite serious since one cannot rely on the perturbative calculations of 
the quark-quark interactions since the color quark current is not conserved.  
The only way out of this dilemma may be to calculate the quark-quark interactions 
with the total Hamiltonian in a non-perturbative fashion since the total 
Hamiltonian is always gauge invariant. At the present stage, we do not know 
which of the shape of the quark-quark inteactions may emerge from the non-perturbative 
treatment. We note that the MIT bag model assumes that the quark-quark interactions 
can be ignored and quarks are free inside hadrons and that the confinement must be 
realized by the boundary condition at the bag surface. 

Here, we also discuss an old but new physical quantity, the conservation 
of the  Noether current in QCD \cite{fay}. As long as a physical quantity like  charge 
is gauge invariant and is conserved, then there is a good chance that it may be observed. 
Here, we discuss a conserved current which is composed of one quark and 
one gluon. This is the only conserved color current of QCD which must play an important 
role for QCD dynamics. Since this current is connected to the  $QG-$state, this state 
can be a physical observable since it is gauge independent and has a conserved charge. 
Therefore, this  $QG-$state has  a finite possibility of existence in nature even though 
it has a color charge and a fractional electric charge. Up to now, there is no observation 
of the fractional charge \cite{perl,gold}, but it does not mean that the object cannot 
be a physical observable, at least, theoretically.

\section{ QCD with $SU(N)$ colors}
The properties of QCD are well discussed in field theory textbooks, but we 
review gauge dependent parts in order to make later discussions easier to understand. 
In particular, we stress that the conservation law of the color current is the only 
conserved current in connection with the gauge fields in the theory. It is a well known 
fact that, in any field theory models  without exception, a conserved charge has been 
playing a dominant role for understanding its dynamics. 
In this sense, it is very strange that the conserved color current has not been 
carefully considered up to now when QCD dynamics is investigated and evaluated, 
and we believe it should be definitely necessary to take into account the current 
conservation  in some way or the other in order to obtain any important information 
of physical observables in  hadron dynamics from QCD.

\subsection{Lagrangian density of QCD}
The Lagrangian density of QCD with $SU(N)$ colors is described as 
$$ {\cal L}= 
\bar{\psi}(i\gamma^{\mu}\partial_{\mu}-g\gamma^{\mu}A_{\mu} -m_0)\psi 
-\frac{1}{2}{\rm Tr}\{ G_{\mu\nu}G^{\mu\nu} \},
 \eqno{(2.1)} $$
where $m_0$ denotes the quark mass, and the field strength $G_{\mu\nu}$ is written as
$$ G_{\mu\nu} = \partial_{\mu}A_{\nu}-\partial_{\nu}A_{\mu}+ig[A_{\mu},A_{\nu}]  
\eqno{(2.2)}$$
$$
A_{\mu} = A^{a}_{\mu}T^{a} \equiv \sum_{a=1}^{n}A^{a}_{\mu}T^{a}  \eqno{(2.3)} $$
where $T^a $ denotes the generator of $SU(N)$ group 
and $n$ is related to $N$ by $n=N^2-1$. 
Here, $T^a $ satisfies the following commutation relations
$$ [ T^{a},T^{b} ]= iC^{abc} T^{c} $$
where $C^{abc}$ denotes the structure constant of the group generators. 

In eq.(2.1), Tr \{ \}  means the trace of the group generators of $SU(N)$, 
and we make use of 
the following identity as a normalization of the representation of the generators
$$ {\rm Tr}\{ T^aT^b\} ={1\over 2} \delta_{ab} .  \eqno{(2.4)}$$ 
Therefore, the last term of eq.(2.1) can be rewritten as 
$$ -\frac{1}{2}{\rm Tr}\{G_{\mu\nu}G^{\mu\nu}\} = -{1\over 4}G_{\mu\nu}^aG^{a,\mu\nu} $$
where $G_{\mu\nu}^a$ is described as
$$ G_{\mu\nu}^a = \partial_{\mu}A_{\nu}^a-\partial_{\nu}A_{\mu}^a-
gC_{abc} A_{\mu}^bA_{\nu}^c . \eqno{(2.5)} $$

\subsection{Infinitesimal gauge transformation} 
The QCD Lagrangian density is invariant under the following infinitesimal local 
gauge transformation with $\chi=\chi^a T^a $
$$ \psi'=(1-ig\chi ) \psi =(1-igT^a\chi^a ) \psi 
 \eqno{(2.6a)} $$
$$ {A'}_\mu =A_\mu +ig[A_\mu,\chi] +\partial_\mu \chi   \eqno{(2.6b)}$$
where $\chi$ is infinitesimally small. 
By defining 
$$ D_\mu =\partial_\mu +igT^aA^a_\mu  \eqno{(2.7)} $$
we can prove  
$$ \bar{\psi'}i\gamma^{\mu}{D'}_{\mu}\psi' 
=\bar{\psi}i\gamma^{\mu}D_{\mu}\psi \eqno{(2.8a)}  $$
$$ {G'}_{\mu\nu}=(1-igT^a\chi^a )G_{\mu\nu}
(1+igT^a\chi^a ) . \eqno{(2.8b)}  $$
Therefore, we obtain
$$  {\rm Tr} \{ {G'}_{\mu\nu}{G'}^{\mu\nu} \} = {\rm Tr} \{(1-igT^a\chi^a ) 
{G}_{\mu\nu}{G}^{\mu\nu}(1+igT^a\chi^a ) \} $$
$$ =   {\rm Tr} \{ {G}_{\mu\nu}{G}^{\mu\nu} \} . \eqno{(2.9)}  $$
This completes the proof of the local gauge invariance of eq.(2.1).

\section{Noether current in QCD} 
The QCD Lagrangian density is invariant under the following infinitesimal global 
gauge transformation
$$ \psi'=(1-igT^a\theta^a ) \psi \eqno{(3.1a)}  $$
$$ {A'}_\nu^a =A_\nu^a -gC^{abc}A^b_\nu\theta^c \eqno{(3.1b)}  $$
where $ \theta^a $ is an infinitesimally small constant. 
In this case, we obtain
$$ \delta {\cal L} = {\cal L}(\psi',\partial_\mu \psi',{A'}_\nu^a, 
\partial_\mu {A'}_\nu^a ) -{\cal L}(\psi,\partial_\mu \psi,{A}_\nu^a, 
\partial_\mu {A}_\nu^a ) =0  . $$
By making use of the equations of motion for $\psi$ and $A^a_{\mu}$, we obtain
$$ \delta {\cal L} = \left[-ig(i\partial_\mu \bar{\psi} \gamma^\mu T^a\psi 
+i\bar{\psi} \gamma^\mu T^a \partial_\mu\psi) \right. $$
$$ \left. +g C^{bca}(\partial_\mu G^{\mu \nu,b}
A_\nu^c +G^{\mu \nu,b}\partial_\mu A_{\nu}^c) 
\right] \theta^a =0 .  \eqno{(3.2)}  $$
Therefore, we easily see that 
$$ \partial_\mu \left(\bar{\psi} \gamma^\mu T^a\psi  +C^{abc }G^{\mu \nu,b} A_\nu^c 
\right) =0 . \eqno{(3.3)}  $$
This means that the Noether current 
$$ I^{\mu,a} \equiv  j^{\mu,a} +C^{abc }G^{\mu \nu,b} A_\nu^c  \eqno{(3.4)} $$
is indeed conserved. That is, 
$$ \partial_\mu I^{\mu,a} =0 \eqno{(3.5)}  $$
where the quark color current $j_\mu^a $ is defined as 
$$ j_\mu^a =\bar{\psi}\gamma_{\mu}T^a\psi . \eqno{(3.6)}  $$
Thus, the quark color current alone is not conserved, and therefore 
there is no conservation of the quark color charge. This is consistent with the fact 
that the color current of quarks is not a gauge invariant quantity. 

\section{Gauge non-invariance of interaction Lagrangian  }
The interaction Lagrangian density of QCD that involves 
quark color currents is written as 
$$ {\cal L}_I= -g j_\mu^a A^{\mu,a} . \eqno{(4.1)} $$
Now, the interaction Lagrangian density $ {\cal L}_I$ is not gauge invariant, 
and therefore if one wishes to make any perturbation calculations involving 
the quark color currents, then one should check it in advance 
whether one can make the gauge invariant quark-quark interactions or not. 

The interaction Lagrangian density is transformed into a new shape under 
the gauge transformation 
$$ {\cal L}_I= -g j_\mu^a (A^{\mu,a}+   \partial^\mu \chi^a  )  \eqno{(4.2)}  $$ 
where the second term is a gauge dependent term. In the same way as QED case 
which will be briefly reviewed in the next section, 
we can rewrite the second term by making use of the conserved current as
$$  -g j_\mu^a \partial^\mu \chi^a = -g  \partial^\mu (j^a_\mu \chi^a)  
+ gC^{abc}\chi^a \partial^\mu G^b_{\mu \nu} A^{\nu,c} . \eqno{(4.3)}  $$ 
The first term is a total divergence and thus does not contribute 
to perturbative calculations. However, there is no way to eliminate the second term 
which is a gauge dependent term. 

Therefore, one sees that one cannot make any  simple-minded perturbative calculations 
of quark-quark interactions in QCD, contrary to the QED case where the electron-electron 
interaction is well defined and calculated as we will see below. 
This means that one cannot define any potential between quarks, and this is 
of course consistent with the picture that the color charge of quarks are 
not gauge invariant, and therefore it is time dependent.  

\section {Gauge invariance of electron-photon interaction in QED}
In QED, the similar situation indeed happens \cite{q2,gross}. That is, the electron-photon 
interaction is, at a glance, gauge dependent since it transforms as, 
$$ {\cal L}_I =- e j_\mu A^{\mu}- e j_\mu \partial^\mu \chi .  \eqno{(5.1)}  $$
However, eq. (5.1) can be rewritten as 
$$ {\cal L}_I =- e j_\mu A^{\mu}- e  \partial^\mu (j_\mu\chi)
+e(\partial^\mu j_\mu )\chi   \eqno{(5.2)}  $$
and one sees easily that the second term in the right hand side is a total divergence and 
thus does not contribute to perturbative evaluations. The third term 
vanishes as long as the current conservation holds, 
$$ \partial^\mu j_\mu =0 .  \eqno{(5.3)}  $$
Therefore, it is clear that the electron-photon interaction is gauge 
independent, and the potential between electrons is a well defined quantity 
in QED under the condition that the vector current is conserved. 
Therefore, one can well calculate electron-electron potentials in the  perturbation 
theory, which are Coulomb potential or some others like hyperfine interactions 

\section{Two choices of gauge fixings}

Now, we present the evaluation of the quark-quark interactions when there is 
no conservation of the quark current. Here, we suppress both 
the color degree and the explicit gluon-gluon vertex in the calculations 
since they are not relevant in the present discussions. 

Since there is no conservation of the quark current, the equation of motion becomes
$$ \partial_\mu F^{\mu \nu} = g (j^\nu+X^\nu ) \eqno{(6.1)} $$
where the extra current $X^\nu$ is introduced by hand here, but, in QCD, 
this corresponds to the gluon currents. 
Therefore, we have $\partial_\mu (j^\mu+X^\mu) =0$, but $\partial_\mu j^\mu \not=0 $. 
In this case, we can easily calculate  the Coulomb interaction for 
the different gauge fixings, the Coulomb gauge $\nabla \cdot \bm{A}=0$ and 
the temporal gauge $A_0 =0$, 
and show that the different choices of the gauge fixing give indeed different 
Coulomb interactions between quarks. This is consistent with 
the observation in section IV. 

\subsection{Coulomb gauge}
The Hamiltonian of quarks interacting with the gauge field can be written 
$$ {\cal H} = \bar{\psi}  \left(-i \bm{\gamma}\cdot \nabla +m  \right)\psi - g \bm{j}\cdot\bm{A}  +gj_0A_0 $$
$$ +{1\over 2}  \left[\dot{\bm{A}}^2-(\nabla A_0)^2  
+ \bm{B}^2 \right] . 
\eqno{(6.2)}  $$
We take the Coulomb gauge 
$$ \nabla \cdot \bm{A} =0 . \eqno{(6.3)}  $$
Since the quark current is not conserved ( $\partial_\mu j^\mu \not= 0$ ), 
the equation of motion for the gauge field $A_0$ becomes from eq.(6.1)
$$ \nabla^2 A_0 = -g(j_0+X_0) . \eqno{(6.4)} $$ 
This is just a constraint which can be easily solved, and we obtain 
$$ A_0 (r)= {g\over 4\pi} \int {\left(j_0(\bm{r}' )+X_0(\bm{r}' )\right) d^3r' 
\over{|\bm{r}'-\bm{r}| }} . 
\eqno{(6.5)}  $$
Now, we can make use of the following relation
$$ {1\over 2}\int (\nabla A_0)^2 d^3r =-{1\over 2}\int (\nabla^2 A_0) A_0 d^3r  $$
$$={g^2\over 8\pi} \int {\left(j_0(\bm{r}' )+X_0(\bm{r}' )\right)
\left(j_0(\bm{r} )+X_0(\bm{r} )\right)d^3rd^3r' \over{|\bm{r}'-\bm{r}| }} 
\eqno{(6.6)}  $$
where the surface integrals are set to zero. 
Therefore, the Coulomb interaction becomes
$$ H_C= \int gj_0(\bm{r} )A_0(\bm{r} )d^3r -{1\over 2}\int (\nabla A_0)^2 d^3r $$
$$ ={g^2\over 8\pi} \int { \left(j_0(\bm{r}' )-X_0(\bm{r}' )\right)
\left(j_0(\bm{r} )+X_0(\bm{r} )\right)d^3rd^3r' 
\over{|\bm{r}'-\bm{r}| }} \eqno{(6.7)}  $$
It is clear that if we set $X_0=0$, then we can recover the normal Coulomb interaction. 

\subsection{Temporal gauge}

Now, we take the $A_0=0$ and therefore the equation of motion for the gauge field beomes 
$$ \nabla \cdot {\partial \bm{A}\over {\partial t}} =-gj_0 . \eqno{ (6.8)} $$ 
Here, there is still a gauge freedom left. Namely, 
$\bm{ E}=-{\partial \bm{A}\over {\partial t}} $ and $\bm{B}=\nabla \times \bm{A} $ 
are invariant under the following transformation
$$ \bm{A}\rightarrow \bm{A}+\nabla \chi(\bm{r}) \eqno{ (6.9)}  $$
where $\chi(\bm{r})$ depends only on the coordinate $\bm r$. 

Therefore, we can write the vector field $\bm{A}$ as
$$ \bm{A}=\bm{A}_T+\nabla \xi, \ \ \ \ \ {\rm with} \ \ \ \  \nabla \cdot \bm{A}_T=0. 
 \eqno{ (6.10)} $$
In this case, the equation of motion for the gauge field becomes
$$ \nabla^2 \phi_0 =-g(j_0+X_0), \ \ \ \ \ \ {\rm with} \ \ \ \   
\phi_0 \equiv \dot{\xi}. \eqno{ (6.11)}  $$
Therefore, ${1\over 2} \dot{\bm{A}}^2 $ term in the Hamiltonian can be written with 
$\bm{E}_T=-{\partial \bm{A}_T\over {\partial t}} $ 
where $\bm{E}_T$ is the transverse electric field 
$$ {1\over 2} \int \dot{\bm{A}}^2 d^3r = {1\over 2} \int \bm{E}_T^2 d^3r + \int \bm{E}_T 
\cdot \nabla \phi_0 d^3r +{1\over 2} \int (\nabla \phi_0 )^2 d^3r  .  \eqno{ (6.12)} $$
The second term in the right hand side vanishes and the third term is just the same 
as the normal Coulomb interaction. Therefore, the Coulomb interaction 
with the temporal gauge becomes 
$$ H_C= {g^2\over 8\pi} \int {\left(j_0(\bm{r}' )+X_0(\bm{r}' )\right)
\left(j_0(\bm{r} )+X_0(\bm{r} )\right)  d^3rd^3r' 
\over{|\bm{r}'-\bm{r}| }} \eqno{(6.13)}  $$
which is different from the Coulomb interaction from the Coulomb gauge fixing. 
Clearly again, if the extra-current $X_0$ is switched off, then we can recover 
the normal Coulomb interaction which, of course, agrees with the one that is 
obtained from the Coulomb gauge fixing.

Therefore, we see that the different choices of the gauge fixing give different 
Coulomb interactions when the current is not conserved. This confirms that 
the gauge invariance of the interaction term ${\cal L}_I =- g j^a_\mu A^{\mu,a}$ 
cannot be recovered for the nonabelian 
gauge field theory like QCD since the quark currents alone are not conserved. 
This means that the quark-quark interactions have some difficulties when 
one starts from the perturbation theory. 

It is unfortunate that one gluon exchange potential which has been very popular 
up to now does not make sense if one starts to calculate it from 
the perturbation theory. We should find out some alternative expressions for 
interactions between quarks and gluons.

\section{  $QG-$state} 
From eqs.(3.4) and (3.5), one sees that the color octet vector current of one quark and 
one gluon state  ( $QG-$state ) is conserved. Therefore, the color 
charge $Q^a_I$ of the $QG-$state which is defined as
$$ Q^a_I = \int I^{0,a} d^3x  \eqno{(7.1)}  $$
is indeed a conserved quantity even though it is not a color singlet. 
In addition, the charge $Q^a_I$ of  $QG-$state is a gauge invariant quantity. 

Here, one may ask as to whether this $QG-$state 
can be observables or not. The answer is yes, that is, it may physically 
exist since it is gauge independent and a conserved quantity. 
At least, there is no reason to say that it cannot exist theoretically. 

One may then ask whether this is a bound state or not. But we believe 
it is not an important question whether it may be a bound state or 
something else. It is for sure that, once it were made, then there is 
no way to decay since quark or gluon alone cannot exist by themselves. 
The only question, though rather serious, is that, in which way 
the  $QG-$state can be created. Or in other words, what should be a probability 
of producing this object from any kind of hadronic or leptonic reactions ? 

For the above statement, one may claim that there is no fractional electric 
charge observed in nature \cite{perl,gold}, and therefore it should not exist. 
This may be right, but at the present stage, we cannot claim more 
than what it is like. 

Since we prove that the one gluon exchange potential between quarks is not a well defined 
quantity, we should ask what is the potential between two nucleons. To this question, 
we can answer to some extent. 
Firstly, the most important nucleon-nucleon potential  must be given by the one pion 
exchange process. This is quite reasonable since pion which is a color singlet object 
can propagate between nucleons and should indeed produce a one pion exchange potential 
which is a favorable object from experiments. In addition to the boson exchange 
interactions, the exchange of QG states between nucleons may give rise to 
a nucleon-nucleon potential which should be somewhat similar to the exchange force 
in atomic electrons. This may cause  a repulsive interaction between nucleons 
and it should be rather short range interactions. However, this mechanism of the 
QG exchange force should be a future problem to be solved.

\section{ Conclusions}

We have examined the gauge invariance of the quark-gluon interaction in QCD. 
In most of field theory text books, people have carefully discussed the gauge dependence 
of the electron-photon interaction in QED and confirmed that the apparent gauge 
dependence of the interaction can be well eliminated. However, for QCD, the gauge 
invariance of the quark gluon interaction in the context of the perturbation theory 
has not been examined carefully in any of field theory textbooks. This may be because 
the QCD case was treated just in the same as the QED case. Obviously, there is a serious 
difference between QCD and QED, which is essentially due to the nonabelian character 
of the gauge fields. Here, we have shown that the quark-gluon interaction in QCD is 
always gauge dependent and there is no way to eliminate the gauge dependence 
in the quark-gluon interaction. This means that one cannot define the quark-quark 
potential,  at least, in a perturbative sense. 
Therefore, it should be meaningless to discuss the shape of the quark-quark potential 
in perturbative calculations even though one can treat a heavy quark 
in the non-relativistic kinematics in which the concept of potential itself 
is well defined. 

However, we have not presented any alternative for the quark interactions 
with gluons, and this is a future problem to be solved. 

Further, we point out that there is a conserved quantity which is 
composed out of one quark and one gluon $QG-$state, and this is gauge invariant 
and has a conserved charge. The possibility of its existence in nature 
should be carefully examined.

\end{document}